\documentclass[a4paper,twocolumn,11pt,accepted=2025-06-19]{quantumarticle}
\pdfoutput=1
\usepackage[utf8]{inputenc}
\usepackage[english]{babel}
\usepackage[T1]{fontenc}
\usepackage{amsmath}
\usepackage{hyperref}
\usepackage{amssymb}
\usepackage[numbers]{natbib}
\usepackage{subcaption}

\usepackage{xcolor}
\definecolor{Pank}{rgb}{1.0,0.1,0.5}
\usepackage{cancel}

\newcommand{\change}[1]{\color{black} #1 \color{black}}

\usepackage{tikz}
\usepackage{lipsum}

\newtheorem{defn}{Definition}[section]

\begin{document}

\title{A little bit of self-correction}

\author{Michael Kastoryano}
\email{mika@di.ku.dk}
\affiliation{Department of Computer Science, University of Copenhagen, Denmark}
\affiliation{AWS Center for Quantum Computing, Pasadena, CA}
%\orcid{0000-0002-2445-2701}
\author{Lasse B. Kristensen}
\affiliation{Department of Computer Science, University of Copenhagen, Denmark}
\orcid{0000-0003-0290-4698}
%\thanks{You can use the \texttt{\textbackslash{}email}, \texttt{\textbackslash{}homepage}, and \texttt{\textbackslash{}thanks} commands to add additional information for the preceding \texttt{\textbackslash{}author}. If applicable, this can also be used to indicate that a work has previously been published in conference proceedings.}
\author{Chi-Fang Chen}
\affiliation{Institute for Quantum Information and Matter, California Institute of Technology, Pasadena, CA, USA}
\author{András Gilyén}
\affiliation{HUN-REN Alfréd Rényi Institute of Mathematics, Budapest, Hungary}
\maketitle

\begin{abstract}
We investigate the emergence of stable subspaces in the low-temperature quantum thermal dynamics of finite spin chains.  Our analysis reveals the existence of effective decoherence-free qudit subspaces, persisting for timescales exponential in $\beta$.  
Surprisingly, the appearance of  metastable subspaces is not directly related to the entanglement structure of the ground state(s). Rather, 
they arise from symmetry relations in low-lying excited states. Despite their stability within a 'phase', practical realization of stable qubits is hindered by susceptibility to symmetry-breaking  perturbations. This work highlights that there can be non-trivial quantum behavior in the thermal dynamics of noncommuting many body models, and opens the door to more extensive studies of self-correction in such systems.  
\end{abstract}

The pursuit of a self-correcting quantum memory, while still largely elusive, has been an exceptionally fertile ground for identifying new approaches to robustly process quantum information \cite{brown2016quantum, kitaev2003fault, bombin2015gauge, bacon2006operator, haah2011local, kastoryano2016quantum, bombin2015single,nathan2024self}.  Ideally, a self-correcting quantum memory would be a system that inherently stabilizes itself against thermal, stochastic and coherent noise.  A less stringent, yet still valuable, form involves a qubit subspace that is robust against thermal fluctuations but may not withstand arbitrary non-thermal perturbations. At present, most of our understanding of thermal stability is based on commuting models \cite{alicki2009thermalization,alicki2010thermal}, mostly because, until recently, no reliable mathematical model existed for studying the thermalization of noncommuting many-body systems.

The interplay between quantum mechanics and statistical mechanics becomes particularly intriguing at low temperatures. While the high-temperature behavior of non-commuting systems tends to mimic classical Gibbs states \cite{bakshi2024high,rouze2024efficient}, low temperatures reveal a richer structure. The characteristics of the low-lying quantum states, particularly entanglement, intertwine with the statistical features of the Gibbs distribution. When the temperature is lower than the spectral gap, we expect the Gibbs state to approximate the ground state, suggesting that the quantum thermal dynamics should mirror the properties of the ground state. This leads to the hypothesis that the mixing time---the timescale over which the system reaches thermal equilibrium---might diverge at a quantum phase transition at low temperatures. Such a hypothesis  would be consistent with the statics-dynamics connection that is well studied for classical spin systems \cite{martinelli1999lectures} and commuting quantum spin systems \cite{kastoryano2016quantum,capel2021modified}.

\begin{figure}[ht]
    \centering
    \includegraphics[width=\linewidth]{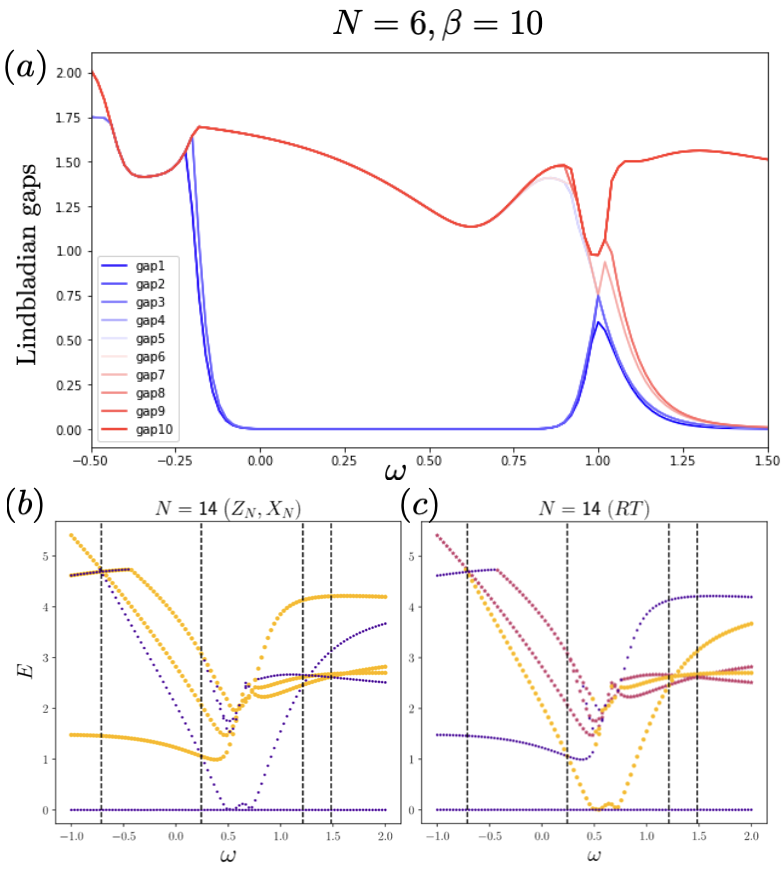}
    \caption{(a)  Numerical simulation of the spectral gaps of the Lindbladian in Eqn. (\ref{eqn:lind}) for the $J_1$-$J_2$ model, with $N=6$ sites, and $\beta=10$. The zero eigenvalue of $\mathcal{L}$ is omitted.  $\beta$ is in the low temperature regime ($\beta\gg {\rm gap}\change{{}^{-1}}(H)$) everywhere except around the Majumdar-Ghosh point ($\omega=.5$), where the Hamiltonian gap closes. Note that the operator $\mathcal{L}$ has real spectrum because of detailed balance \cite{QGS1}. The region $0\leq\omega\leq .8$ has four nearly degenerate eigenvalues, while the region $\omega\geq 1.5$ has nine, reflecting qubit and qutrit metastable subspaces, respectively. 
    (b) Low lying spectrum of $H$; i.e. $e_k-e_0$ for $k=1, ..., 12$. The excited eigenstates are color coded to indicate which symmetry sector they belong to.  The eigenstates $|e_k\rangle$, $k\neq 0$, are plotted in blue if $\langle e_k|Z_N|e_k\rangle = \langle e_0|Z_N|e_0\rangle$ and $\langle e_k|X_N|e_k\rangle = \langle e_0|X_N|e_0\rangle$, and plotted in yellow otherwise. (c)  The eigenstates are plotted in blue if $\langle e_k|RT|e_k\rangle = \langle e_0|RT|e_0\rangle$, in yellow if  $\langle e_k|RT|e_k\rangle \neq \langle e_0|RT|e_0\rangle$, and in pink if $|e_k\rangle$ is not a simultaneous  eigenstate of $R$ and $T$. \change{Numerics were performed using the QuTiP library~\cite{johansson2012}.}}
    \label{fig:one}
\end{figure}

 Building on the new quasi-local noncommutative Gibbs samplers constructed in Refs.  \cite{QGS1,QGS2,QGS3,QGS4}, we investigate the low-temperature quantum thermal dynamics of a specific class of one-parameter quantum spin chains: the $J_1-J_2$ chain. Spin chains are notable as they lack a thermal phase transitions, yet they undergo quantum phase transitions \cite{sachdev1999quantum}. Our analysis reveals that:
(i) There exist qudit subspaces that remain decoherence-free for timescales on the order of 
$\Omega(N^{-1} e^{\beta c})$, 
where $\beta$ is the inverse temperature, $c$ is a constant related to gaps in the low lying spectrum of the Hamiltonian, and $N$ is the chain length. (ii) The existence of  metastable subspaces does not require ground state degeneracy of the model. Rather, it originates from symmetry constraints in the low lying spectrum, and depends on the system-bath coupling operators. (iii) The 'phase diagram' of these stable subspaces does not necessarily mirror the ground state quantum phase diagram. Instead, the metastable subspaces can appear and disappear at (avoided-)crossings of the low-lying excited states of the model (see Fig. \ref{fig:one}). 

Despite the intriguing nature of these findings, we temper expectations as for their practical utility in building noise resilient qubits. The primary limitation stems from the protection mechanism's reliance on  discrete symmetries. Such symmetries are vulnerable to local perturbations, which can easily disrupt the stability afforded by these decoherence-free subspaces. Consequently, while the stable subspaces reveal interesting new physics in the thermal dynamics of quantum systems, they may not translate into practical, fault-tolerant quantum memories.

This work is  significant for several reasons. First, it provides initial evidence of non-trivial quantum behavior of thermalization. This marks a departure from the classical understanding, showcasing how quantum effects can alter thermal dynamics in unexpected ways. Second, it highlights the critical influence of the choice of elementary jump operators on the system's dynamics. This observation suggests that subtle details in the model's implementation can have profound impacts on its behavior. Finally, the presence of approximate decoherence-free subspaces within the thermal dynamics of spin systems suggests that such phenomena might be more common than previously thought, opening new avenues for exploring the interplay between quantum coherence and thermal noise.

\section{Metastability}

While ground states of commuting Hamiltonians always have finite correlation length,  noncommuting quantum many-body systems can have extensive ground state correlations, allowing for various forms of topological and symmetry-driven quantum phase transitions \cite{sachdev1999quantum,zeng2019quantum}. While the quantum thermal dynamics of commuting systems is fairly well understood \cite{kastoryano2016quantum,bardet2023rapid}, the quantum thermal dynamics of noncommuting many-body systems is essentially uncharted territory. This is due in large part to the lack of a go-to quantum thermal dynamics algorithm. Recently, a subset of the authors introduced a (quasi-)local quantum Lindbladian that satisfies detailed balance with respect to a noncommuting quantum Gibbs state \cite{QGS2}, hence allowing for the analytic and numerical analysis of quantum Gibbs sampling. This new construction will be the starting point for our analysis.

We consider a quantum system described by a local many-body Hamiltonian $H$, on a regular lattice $\Lambda$. 
The thermal Markovian dynamics elaborated in Refs. \cite{QGS1,QGS2,QGS3,QGS4} is defined by a Lindbladian (the generator of a quantum dynamical semigroup)
\begin{equation}
    \mathcal{L}(\rho)= -i[Q,\rho]+\sum_a L_a \rho L_a^\dag -\frac{1}{2}(L_a^\dag L_a\rho+\rho L_a^\dag L_a).\label{eqn:lind}
\end{equation}
$L_a$ are the quantum  jumps, or Lindblad operators, and are given explicitly by 
\begin{equation}\label{eqn:jumps}
    L_a = \int_{-\infty}^\infty g(t) e^{-itH} S_a e^{i t H} dt, 
\end{equation}
where $\{S_a\}$ are a complete set of elemental jump operators that guarantee ergodicity of the dynamics. In the case of a quantum spin model on a 1D lattice (a ring), it is customary to choose $\{S_a\}$ to be the set of local Pauli operators: $S_a\in \{X_n,Y_n,Z_n\}_{n\in \Lambda}$. $g(t) = \int_{-\infty}^\infty \sqrt{\gamma(\omega)}e^{it\omega }d\omega$, is a damping term, and  $\gamma(\omega)$ is related to the bath auto-correlation function, and incorporates the system bath coupling strength. For concreteness, we assume that $\gamma(\omega)\propto 1/(1+e^{\beta \omega})$ for a specific temperature $\beta$. This choice is reminiscent of an Ohmic bath \cite{gardiner2004quantum}, but all of our conclusions hold more generally, as long as $\gamma(-\omega)=e^{\beta \omega}\gamma(\omega)$. We choose to work with the specific form of jump operators in Eqn. (\ref{eqn:jumps}) for concreteness, however there is a wider class of constructions which lead to the same conclusions \cite{QGS3,QGS4}.

$Q$ is a Hermitian matrix, playing the role of the system Hamiltonian. The two natural candidates for $Q$ are (i) $Q\propto H$, the Hamiltonian of the system, and (ii) 
\begin{equation}\label{eqn:DBQ}
    Q=\frac{i}{2}\sum_{ij}\tanh\left(\frac{\beta(e_i-e_j)}{4}\right)\langle e_i|\sum_a L_a^\dag L_a|e_j\rangle |e_i\rangle\langle e_j|,
\end{equation}
written out in the eigenbasis of $H=\sum_j e_j |e_j\rangle\langle e_j|$. The latter choice guarantees that the stationary state of the dissipative dynamics is the Gibbs state; i.e. $\mathcal{L}(\rho_\beta)=0$ with $\rho_\beta\propto e^{-\beta H}$. The specific form in Eqn. (\ref{eqn:DBQ}) is necessary to enforce the quantum detailed balance condition \cite{QGS2,QGS3,QGS4}---the key mathematical relationship governing both classical and quantum Gibbs samplers.  The first choice $Q=H$ is closer to  the natural dynamics resulting from a Markovian weak-coupling limit, as shown in Ref. \cite{nathan2020universal}. However, it does not lead to a stationary state that is globally Gibbs at any temperature. We will see that at low enough temperatures both choices lead to metastable quantum subspaces, though $Q=H$ induces coherent dynamics in the metastable subspace.

It is worth pausing at this point to discuss how important the choice of bath couplers $\{S_a\}$ is for the thermalization process. By analogy with the classical Metropolis algorithm, it is generally believed that as long as the operators $\{S_a\}$ are local on the lattice, and span the full algebra of the space, then the dynamics will not depend sensitively on the specific choice of couplers $\{S_a\}$. \change{This intuition builds upon a deep correspondence from Glauber dynamics of classical lattice spin systems, which shows a strong correspondence between static properties of the Gibbs states and the mixing time of the dynamics \cite{martinelli1999lectures}, at least in the case of finite-dimensional lattice systems \cite{berger2005}.}
We show, on the contrary, that the choice of couplers plays a crucial role in the quantum case in the large $\beta$ limit, when the statics is dominated by the ground state.

%%%%%%%%%%%%%%%%%%%%%%%%%%%%%%%%%%%%%%%%%%%%%%%%%%%

\paragraph{Decoherence-free subspaces}
%We start by recalling the definition of a decoherence free subspace of a Lindbladian. 
We say that a Lindbladian admits a decoherence-free subspace (DFS) $\mathcal{D}$ if for any state $\psi\in\mathcal{D}$, $\mathcal{L}(\psi)=0$ \cite{knill2000theory,zanardi2000stabilizing}.
Equivalently \cite{kraus2008preparation}, $\mathcal{D}$ is a DFS of $\mathcal{L}$ if    $L_a |\psi\rangle = 0$ and $Q |\psi\rangle \propto |\psi\rangle$,
for all Kraus operators $L_a$, and any pure state $|\psi\rangle\in\mathcal{D}$. We will consider an $\epsilon$-approximate version of the DFS ($\epsilon$-DFS):
\begin{eqnarray}
    ||L_a |\psi\rangle|| &\leq& \epsilon||L_a||\\
    ||Q|\psi\rangle|| &\leq& \change{\epsilon^2 ||Q||}.
\end{eqnarray}
In general, the Hamiltonian only needs  $|\psi\rangle$ to be an approximate eigenstate for the subspace to be $\epsilon$-DFS, but for our purpose, the more restrictive definition is sufficient. As mentioned above, this requirement can be further relaxed to allow for unitary dynamics within the subspace~\cite{lidar2003}, as occurs in the case $Q=H$.
\begin{defn}[Blind qubit subspace]
Consider a Hamiltonian $H=\sum_j e_j |e_j\rangle\langle e_j|$, with its eigenvalues in increasing order: $e_0\leq e_1\leq ...$, and a set of elemental jump operators $\{S_a\}$. If there exists an eigenvector $|e_k\rangle$, $k>0$  such that $\langle e_j|S_a|e_k\rangle=0$ for all $j\leq k$, and $\langle e_0|S_a|e_0\rangle=0$, then $\mathcal{S}={\rm span}(|e_0\rangle,|e_k\rangle)$ is a \textit{blind qubit subspace} of $\{S_a\}$. 
\end{defn}
The extension to qudit subspaces is straightforward.
We now show that a blind subspace of $\{S_a\}$ is metastable. %an $\epsilon$-DFS of $\mathcal{L}$ with $\log(\epsilon)\propto -\beta$.
Let $\mathcal{S}$ be a blind subspace of $\{S_a\}$, and consider $\psi\in\mathcal{S}$. We time evolve this state with respect to the thermal Lindbladian in Eqn. (\ref{eqn:lind}) for a time $t$. Then, note that \cite{chesi2010thermodynamic}
\begin{equation}\label{eqn:meta}
    ||\psi(t) - \psi||_1 = ||\int_0^t  \dot{\psi}(s) ds||_1\leq t||\mathcal{L}(\psi)||_1.
\end{equation}
Writing the Lindbladian as $\mathcal{L}(\rho)=T(\rho)+(\kappa \rho+\rho \kappa^\dag)$, with $T(\rho)=\sum_a L_a\rho L_a^\dag$ and $\kappa = - \frac{1}{2} \sum_a L_a^\dag L_a \color{black} -i Q$ we get
\begin{equation}
    ||\mathcal{L}(\psi)||_1\leq 2||\kappa|\psi\rangle||+||T(\psi)||_1.
\end{equation}
If $\psi$ is in an $\epsilon$-DFS subspace, then $||\kappa |\psi\rangle ||=O(\epsilon N)$, and $||T(\psi)||_1= O(\change{\epsilon^2} N)$, where $N=|\Lambda|$ is the volume, so the dominant contribution will come from $||\kappa|\psi\rangle||$ term.
%\change{Thus, $\epsilon$-DFS'es are characterized by metastability.}

Let us now look at the specifics of the thermal Lindbladian. 
Consider  the action of $L_a$ on an eigenstate of $H$:
\begin{eqnarray}
    L_a |e_j\rangle&=& \int_{-\infty}^\infty g(t) e^{-itH} S_a e^{i t H}|e_j\rangle \,dt\\
    &=& \sum_{k} \int_{-\infty}^\infty g(t) e^{-it (e_k-e_j)} \langle e_k|S_a|e_j\rangle |e_k\rangle \,dt \nonumber\\
    &=&  2 \pi \color{black} \sum_k \sqrt{\gamma(e_k-e_j)} \langle e_k|S_a|e_j\rangle |e_k\rangle.\label{eqn:main_suppression}
\end{eqnarray}

By construction, note that $\gamma(x)\leq 1$, then combined with the property that $\gamma(\omega)=e^{-\beta \omega}\gamma(-\omega)$, we have that $\gamma(\omega)\leq e^{ - \beta \omega}$. Then
\begin{eqnarray}
    ||L_a|e_j\rangle|| &\leq&  2 \pi \sum_ke^{\beta(e_j-e_k)/2}|\langle e_k|S_a|e_j\rangle| \label{eq:jump_op_norm}\\
    &\leq&  2 \pi  \sum_{k=1}^j e^{\beta(e_j-e_k)/2} |\langle e_k|S_a|e_j\rangle| \\
    & \, &  + \; O(e^{-\beta \Delta/2}),
\end{eqnarray}
where  $\Delta=\min\{e_{j+1}-e_j,e_p-e_0\}$ with $\left|e_p\right>$ the lowest-lying state not in the blind subspace, and we assume that $\sum_{k>j} |\langle e_k|S_a|e_j\rangle| e^{-\beta(e_k-e_{j+1})/2}=O(1)$. The latter assumption will be true whenever the bulk of the eigenvalues of $H$ is  sufficiently removed from the extremities of the spectrum, as is typically the case for local lattice models \footnote{See the appendix \ref{app:boltzmann_sum_alt} for a discussion of this assumption in the setting of the $J_1-J_2$ model.}. 
If $|e_0\rangle$ and $|e_j\rangle$ span a \textit{blind qubit subspace} with respect to the elemental jump operators $\{S_a\}$, i.e.
\begin{equation}
    |\langle e_k|S_a|e_j\rangle| =0 
\end{equation}
for all $k\leq j$, then we get that
\begin{equation}
    ||L_a|e_j\rangle|| \leq O(e^{-\beta \Delta/2}).
\end{equation}

The 'Hamiltonian' term $Q$ given by Eqn. (\ref{eqn:DBQ}) can be controlled similarly (see the appendix), leading to an overall scaling of $||\mathcal{L}(\psi)||_1=O(N e^{-\beta \Delta/2})$. Hence, from Eqn. (\ref{eqn:meta}), any state in the blind subspace survives for a time $\Omega(N^{-1}e^{\beta \Delta/2})$. The coefficient $\Delta$ is governed by the gaps above $e_0$ and $e_j$. 

\section{Symmetries}

\begin{figure}[ht]
    \centering
    \includegraphics[width=1\linewidth]{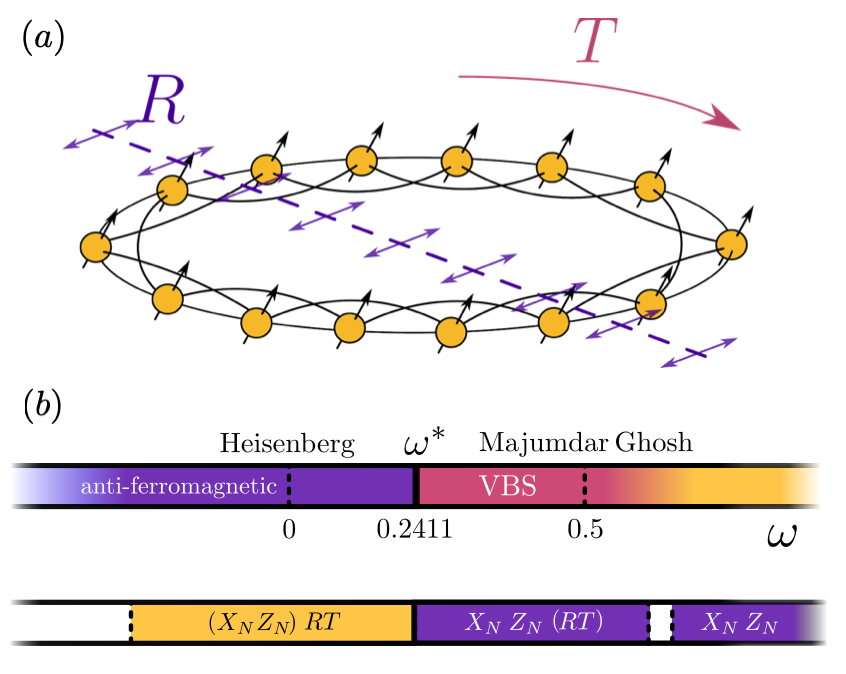} 
    \caption{(a) Illustration of the $J_1-J_2$ spin chain, with reflection $R$ and translation $T$ symmetries. (b) Ground state phase diagram of the model with $\change{\omega}\equiv J_1/J_2$, and low temperature metastability diagram indexed by the protecting symmetries for $N=14$. The locations of the metastability diagram depend on the chain length $N$.}
    \label{fig:two}
\end{figure}

In order to illustrate the existence of metastable subspaces in the thermal dynamics, we will concentrate primarily on  the $J_1-J_2$ model, which exhibits a rich ground state and dynamical symmetry phase diagram (see Fig. \ref{fig:one}). The 1D quantum $J_1$-$J_2$ model describes the behavior of spins on a one-dimensional lattice with competing nearest-neighbor ($J_1$) and next-nearest-neighbor ($J_2$) exchange interactions. The Hamiltonian is given by:
\begin{equation}
H =  \sum_{n=1}^N J_1 \mathbf{S}_n \cdot \mathbf{S}_{n+1} + J_2  \mathbf{S}_n \cdot \mathbf{S}_{n+2} ,
\end{equation}
where  $\mathbf{S}_n=(X_n,Y_n,Z_n)^T$ in the spin-$1/2$ case. We will assume periodic boundary conditions and $N$ even. $J_1$ and $J_2$ are the (real) exchange coupling constants. Without loss of generality, we will fix $J_1=1$ and write $J_2/J_1\equiv \omega$. The system has a gappeless unique ground state when $\omega< \omega^*$, and a gapped phase for $\omega> \omega^*$ with degenerate ground state in the thermodynamic limit. The system undergoes a Kosterlitz-Thouless transition at $\omega^*\approx 0.2411$ \cite{okamoto1992fluid}. The point $\omega=0.5$---the Majumdar-Ghosh point---has a known exact solution. Specifically, the degenerate ground states at $\omega=0.5$ have an exact representation as matrix product states. 
The spin-$1/2$ model enjoys a number of discrete symmetries: total 'charge' $X_N = \bigotimes_n X_n$, total 'phase' $Z_N=\bigotimes_n Z_n$, translation symmetry $T: S^\alpha_n\rightarrow S^\alpha_{n+1}$, and reflection symmetry $R: S^\alpha_n\rightarrow S^\alpha_{L-n+1}$, along any line cutting the ring in two equal-size pieces (see Fig. \ref{fig:two}). 

We start our analysis with the spin-$1/2$ case at the Majumdar-Ghosh point: $\omega=0.5$. At this special point, the model has a degenerate ground subspace spanned by the states:    $|L\rangle = \bigotimes_{e\in E_{\rm even}} |-_e\rangle$ and $|R\rangle = \bigotimes_{o\in E_{\rm odd}} |-_o\rangle,$
where $|-_n\rangle = (|00\rangle_{n,n+1}-|11\rangle)_{n,n+1}/\sqrt{2}$. The tensor product is taken over even edges of the ring for the $|L\rangle$ state and over odd edges for the $|R\rangle$ state. These are known as valence bond solid (VBS) states. The model is believed to remain in the VBS phase all the way down to the phase transition point $\omega^*$. A straightforward calculation shows that the states $|L\rangle,|R\rangle$ are protected against local Pauli errors:
\begin{equation}\label{eqn:LRmg}
    \langle L|P_n|R\rangle =\langle L|P_n|L\rangle = \langle R|P_n|R\rangle=0,
\end{equation}
where $P_n\in\{X_n,Y_n,Z_n\}$.  To see this, note that when $N$ is even, $X_N$, $Z_N$ and $H$ all commute with each other, therefore the eigenvectors of $H$ are also eigenvectors of $X_N$ and $Z_N$ with eigenvalues $\pm 1$. In particular,  $Z_N|L,R\rangle=|L,R\rangle$ and $X_N|L,R\rangle=|L,R\rangle$ for $N/2$ even, and $X_N|L,R\rangle=-|L,R\rangle$ for $N/2$ odd \footnote{$|L,R\rangle$  is shorthand notation to mean $|L\rangle$ or $|R\rangle$.}. Then, for any site $n=1,2,...$, we note that 
\begin{eqnarray}
    \langle L| Z_n|L\rangle &=& \langle L| X_N Z_n X_N|L\rangle\\
    &=& -\langle L| Z_n|L\rangle,
\end{eqnarray}
where the second equality follows from anti-commutation of $X_N$ and $Z_n$. It follows that the expectation value $\langle L| Z_n|L\rangle$ has to be zero. The same argument holds for all combinations of $|L\rangle, |R\rangle$ and all local Pauli operators $\{X_n,Y_n,Z_n\}$. All we need for the argument to hold is for $|L\rangle$ and $|R\rangle$ to have the same eigenvalue under $Z_N$ and $X_N$. It turns out \cite{caspers1982some} that the third eigenstate does not satisfy this property, hence the protection is restricted to a two-dimensional subspace. 
By the arguments above, the subspace spanned by $|L\rangle,|R\rangle$ is a blind subspace with respect to local Paulis, and hence an approximate DFS under this noise. 
 Note that the constraints arising from discrete symmetries are somewhat reminiscent of the selection rules governing allowed and forbidden transitions in nuclear, atomic, and molecular systems \cite{hamermesh2012, bransden2003, campos2021}.% {\color{Pank} [I think the sentence starting here sticks out as a bit odd and could be removed]} The main difference being that in those cases the matrix elements are of couplings to external electromagnetic fields, and the symmetries at play are typically spatial symmetries like inversion and rotation. Note further, that Eqn. \eqref{eqn:LRmg} does not hold if the bath coupling operators contain even products of Paulis.  

Away from the Majumdar-Ghosh point, the model can only be explored numerically \footnote{Some analytic expressions exist for excited states \cite{caspers1982some, arovas1989two}, but there is no guarantee that these states are the first few excited states}.  
In Fig. \ref{fig:one}b, we plot the low lying spectrum of $H$ for $\omega\in [-1,2]$ and $N=14$. We color code each excited state according to  whether it has the same symmetry as the ground state (blue), it has a different symmetry (yellow), or the state is not an eigenstate of the symmetry (pink). We plot this for the $X_N,Z_N$ symmetries and for the $RT$ symmetry.   What we observe in the figure is that for $\omega^*< \omega \lesssim 1.2$, the two lowest states are protected by the $X_N,Z_N$ symmetry. For $\omega \gtrsim 1.5$, the model's first excited states are degenerate, but satisfy the same symmetry conditions as the ground state, hence yielding a qutrit blind subspace. Below $\omega_L\sim -.8$, there is no blind subspace, as all of the symmetry conditions are violated.   

For $ -.8 \lesssim\omega<\omega^*$, the excited states violate the $X_N,Z_N$ symmetry conditions, suggesting no $\epsilon$-DFS. However, a closer inspection shows that a new subspace emerges, protected by both the $X_N,Z_N$ and the $R,T$ symmetries.  Note first that, in general, translations and reflections do not commute, $[R, T] \neq 0$. This means that even though they each commute with the Hamiltonian, we are not guaranteed the existence of a full set of states that are simultaneous eigenstates of all three operators $\{H, T, R\}$. For simplicity, we define our states as eigenstates of $H$ and $T$. This has the benefit that in the cases where a state is also symmetric with respect to a reflection, translation symmetry guarantees that it will respect \emph{all} of the possible reflections of the ring, with identical eigenvalue.  Now, observe that for any given single site Pauli $P_n$, we can choose a reflection $R_n$ such that $R_nP_nR_n=P_{n-1}$. Then, the $TR_n$ transformation yields $T^\dag R_n P_n R_n T = P_n$. The same argument holds for any single-site Pauli operator.  Assuming that low-lying states $\{|e_j\rangle\}$ are eigenstates of both $T$ and reflections, and denoting the eigenvalues as $T|e_j\rangle = \lambda_j^T|e_j\rangle$ and $R_n|e_j\rangle=\lambda_j^{R_n}|e_j\rangle$, we get 
\begin{equation}
    \langle e_j|P_n|e_k\rangle = \lambda_j^{R_n}\lambda_k^{R_n}\bar{\lambda}^T_j\lambda^T_k\langle e_j|P_n|e_k\rangle.
\end{equation}
Hence, $\langle e_j|P_n|e_k\rangle$ must be zero, unless $\lambda_j^{R_n}\lambda_k^{R_n}\bar{\lambda}^T_j\lambda^T_k=1$. Recall, because of translation invariance, $\lambda^{R_n}$ is independent of the site $n$.

Consider now the table of symmetries at the value $\omega=0$ (i.e. the Heisenberg model) for the five lowest eigenvectors:  

\begin{table}[ht]
\centering
\begin{tabular}{c|c|c|c|c}

$\omega=0$ & $\lambda^{X_N}$ & $\lambda^{Z_N}$ & $\lambda^R$ & $\lambda^T$ \\
\hline\hline
$|e_0\rangle$ & $-1$ & $-1$ & $-1$ & $-1$ \\
\hline
$|e_1\rangle$ & $1$ & $1$ & $1$ & $1$ \\
\hline
$|e_2\rangle$ & $-1$ & $1$ & $1$ & $1$ \\
\hline
$|e_3\rangle$ & $1$ & $-1$ & $1$ & $1$ \\
\hline
$|e_4\rangle$ & $-1$ & $-1$ & $-1$ &  $1$ \\

\end{tabular}
\label{tab:example}
\end{table}

From the table of symmetries, which was obtained numerically, and the argument above, it follows that $\langle e_j|P_n|e_4\rangle =0$ for $j=0,1,2,3,4$ at the Heisenberg point. Given that there is a gap to the fifth excited eigenstate, there therefore exists a blind qubit subspace spanned by $|e_0\rangle$ and $|e_4\rangle$. The blind subspace is protected by all of the unitary symmetries of the system\change{, but predominantly by the $R$ and $T$ symmetries.} We expect the state $|e_4\rangle$ to be related to the anti-ferromagnetic ground state via the Bethe ansatz with total magnetization zero, as a sum of string operators, as argued in Ref. \cite{faddeev1984spectrum,doikou1999parity}. The metastable subspace is also found in the Heisenberg XXZ model throughout the paramagnetic region. This is relevant, as physical implementation of next-nearest-neighbor models are much more challenging on present-day hardware than nearest-neighbor models.

\paragraph{Topological stabilizer Hamiltonians.}
Another class of models which very naturally exhibits the conditions for an $\epsilon$-DFS are topological stabilizer models. Indeed, consider the 1D Wen  model \cite{wen2003quantum}
$H_{\rm wen}=\change{\sum_{n=1}^{N}} h_n$, with $h_n = X_nZ_{n+1}Z_{n+2}X_{n+3}$, and again assume periodic boundary conditions. There are two stabilizer/ground states of the model, both satisfying $h_n|\psi\rangle=0$ for all $n$. Then, we note that for any local Pauli operator $P_m$, we can find a stabilizer $h_n$ such that $h_n P_m h_n = -P_m$. This implies that $\langle \psi| P_m |\psi'\rangle=0$, for any $|\psi\rangle,|\psi'\rangle$ in the stabilizer subspace. In fact this even holds of every string of Pauli operators that are not stabilizer-equivalent to the $X_N$ and $Z_N$ operators \footnote{$A$ and $B$ are stabilizer equivalent if there exists a product of stabilizers $\Omega=\prod_\alpha h_{j_\alpha}$ such that \change{$A=\Omega B$}.}; i.e. the logical operators of the code when $N$ is odd. In this case, the protection is due to local symmetries of the stabilizer group.

\subsection{Robustness} The type of robustness afforded by the $\epsilon$-DFSs depends crucially on the symmetries protecting the subspace. Going over the models covered here, the VBS-type phase, protected by global charge $X_N$ and global phase $Z_N$, will be stable to any weak perturbation of the Hamiltonian preserving the symmetries. This includes any term of the form $XX$, $ZZ$ on any two sites, and arbitrary combinations thereof. However, any 'magnetic field' (i.e. single or odd Pauli-weight terms) will \change{in general} break the metastability. \change{One exception is translationally invariant combinations of such perturbations, since the $RT$ symmetries ensure metastability in this phase even when the $X_N$ and $Z_N$ symmetries are broken.} 
The anti-ferromagnetic phase is \change{partly} protected by $X_N$, \change{and} $Z_N$, \change{but only fully protected by} $RT$. Hence, only transitionally invariant perturbations are allowed, which is a much stronger restriction. \change{For more details on the relationship between symmetries and metastability in these phases, see App.~\ref{app:Sensitivity}.} 

The topological models are in turn only stable to multiplicative prefactors in front of the stabilizer terms, as well as perturbations that are in the stabilizer group. It is important to note that even for the toric code, whose ground subspace is topologically protected, non-stabilizer perturbations will kill the $\epsilon$-DFS, leaving only classical information as metastable states of the thermal dynamics.

\paragraph{Stable qubits} It is tempting to ask whether the metastable subspaces can be used to protect and manipulate logical information. At a first glance, the metastability described in this article looks pretty fragile. However, certain features lend themselves to optimism. First, because the effect described here is not very sensitively dependent on the precise form of the thermal dynamics, only on the system bath couplings, it is reasonable to expect that several of the recently proposed cooling protocols \cite{matthies2022programmable,kishony2023gauged,andersen2024thermalization} might reveal the existence of low-temperature metastable states. Furthermore, in these synthetic situations, it might be possible to enforce the symmetries robustly. 
The error correction properties of the VBS phase have been described extensively in Ref. \cite{lloyd2024quasiparticle,nakamura2002order,wang2020classes}, though explicit constructions are only known for the specific Majumdar-Ghosh point. In particular, the logical operators are only known in the large-$N$ limit. It is unclear at this point how to engineer logical operations at other points in the phase diagram.

\section{Discussion}

The presented results show that there can be metastable states in the low-temperature thermal dynamics of non-commuting spin chain models. Importantly, the existence of the metastable subspaces does not require ground state degeneracy. Conversely, ground state degeneracy is not a sufficient condition either. For instance, the transverse field Ising model in the ferromagnetic phase has exponentially small splitting between the ground state and first excited state but does not possess a metastable subspace, since the ground state expectation value of local Pauli operators is non-zero. As a consequence, there is dephasing within the ground subspace, leading to a two-dimensional stationary space of the Lindbladian; in other words, a stable classical bit.  

One important question which we leave open is whether some weaker forms of protection can be seen in these systems as the chain length grows. So far, our results do not really depend upon chain length within a given phase. Rather, the effects discovered here are most pronounced for small system sizes ($N<12$).  In the appendix, we provide some additional analysis of the behavior of metastable subspaces as the chain length increases. 
Finally, we ask whether similar effects can be seen in systems with additional structure; including, but not limited to, fermionic systems, lattice gauge theories, and subsystem codes. \\

\begin{acknowledgments}
We acknowledge support from the AWS Center for Quantum Computing, and from the Carlsberg foundation.  
\end{acknowledgments}

\bibliographystyle{quantum}
\bibliography{ref}

\begin{thebibliography}{10}

\bibitem{brown2016quantum}
Benjamin~J Brown, Daniel Loss, Jiannis~K Pachos, Chris~N Self, and James~R
  Wootton.
\newblock ``Quantum memories at finite temperature''.
\newblock \href{https://dx.doi.org/10.1103/RevModPhys.88.045005}{Reviews of
  Modern Physics {\bf 88}, 045005}~(2016).

\bibitem{kitaev2003fault}
A~Yu Kitaev.
\newblock ``Fault-tolerant quantum computation by anyons''.
\newblock \href{https://dx.doi.org/10.1016/S0003-4916(02)00018-0}{Annals of
  physics {\bf 303}, 2--30}~(2003).

\bibitem{bombin2015gauge}
H{\'e}ctor Bomb{\'\i}n.
\newblock ``Gauge color codes: optimal transversal gates and gauge fixing in
  topological stabilizer codes''.
\newblock \href{https://dx.doi.org/10.1088/1367-2630/17/8/083002}{New Journal
  of Physics {\bf 17}, 083002}~(2015).

\bibitem{bacon2006operator}
Dave Bacon.
\newblock ``Operator quantum error-correcting subsystems for self-correcting
  quantum memories''.
\newblock \href{https://dx.doi.org/10.1103/PhysRevA.73.012340}{Physical Review
  A {\bf 73}, 012340}~(2006).

\bibitem{haah2011local}
Jeongwan Haah.
\newblock ``Local stabilizer codes in three dimensions without string logical
  operators''.
\newblock \href{https://dx.doi.org/10.1103/PhysRevA.83.042330}{Physical Review
  A {\bf 83}, 042330}~(2011).

\bibitem{kastoryano2016quantum}
Michael~J Kastoryano and Fernando~GSL Brandao.
\newblock ``Quantum {G}ibbs samplers: The commuting case''.
\newblock \href{https://dx.doi.org/10.1007/s00220-016-2641-8}{Communications in
  Mathematical Physics {\bf 344}, 915--957}~(2016).

\bibitem{bombin2015single}
H{\'e}ctor Bomb{\'\i}n.
\newblock ``Single-shot fault-tolerant quantum error correction''.
\newblock \href{https://dx.doi.org/10.1103/PhysRevX.5.031043}{Physical Review X
  {\bf 5}, 031043}~(2015).

\bibitem{nathan2024self}
Frederik Nathan, Liam O'Brien, Kyungjoo Noh, Matthew~H Matheny, Arne~L Grimsmo,
  Liang Jiang, and Gil Refael.
\newblock ``Self-correcting {GKP} qubit and gates in a driven-dissipative
  circuit''~(2024).
\newblock  \href{http://arxiv.org/abs/2405.05671}{arXiv:2405.05671}.

\bibitem{alicki2009thermalization}
Robert Alicki, Mark Fannes, and Michal Horodecki.
\newblock ``On thermalization in {K}itaev's 2d model''.
\newblock \href{https://dx.doi.org/10.1088/1751-8113/42/6/065303}{Journal of
  Physics A: Mathematical and Theoretical {\bf 42}, 065303}~(2009).

\bibitem{alicki2010thermal}
Robert Alicki, Michal Horodecki, Pawel Horodecki, and Ryszard Horodecki.
\newblock ``On thermal stability of topological qubit in {K}itaev's 4d model''.
\newblock \href{https://dx.doi.org/10.1142/S1230161210000023}{Open Systems \&
  Information Dynamics {\bf 17}, 1--20}~(2010).

\bibitem{bakshi2024high}
Ainesh Bakshi, Allen Liu, Ankur Moitra, and Ewin Tang.
\newblock ``High-temperature {G}ibbs states are unentangled and efficiently
  preparable''.
\newblock In 2024 IEEE 65th Annual Symposium on Foundations of Computer Science
  (FOCS).
\newblock \href{https://dx.doi.org/10.1109/FOCS61266.2024.00068}{Pages
  1027--1036}.
\newblock IEEE~(2024).

\bibitem{rouze2024efficient}
Cambyse Rouz{\'e}, Daniel~Stilck Fran{\c{c}}a, and {\'A}lvaro~M Alhambra.
\newblock ``Efficient thermalization and universal quantum computing with
  quantum {G}ibbs samplers''~(2024).
\newblock  \href{http://arxiv.org/abs/2403.12691}{arXiv:2403.12691}.

\bibitem{martinelli1999lectures}
Fabio Martinelli.
\newblock ``Lectures on {G}lauber dynamics for discrete spin models''.
\newblock In Lectures on probability theory and statistics.
\newblock \href{https://dx.doi.org/10.1007/978-3-540-48115-7_2}{Pages 93--191}.
\newblock Springer~(1999).

\bibitem{capel2021modified}
Ángela Capel, Cambyse Rouzé, and Daniel Stilck~França.
\newblock ``The modified logarithmic {S}obolev inequality for quantum spin
  systems: classical and commuting nearest neighbour interactions''~(2021)
  \href{http://arxiv.org/abs/2009.11817}{arXiv:2009.11817}.

\bibitem{QGS1}
Chi-Fang Chen, Michael~J Kastoryano, Fernando~GSL Brand{\~a}o, and Andr{\'a}s
  Gily{\'e}n.
\newblock ``Quantum thermal state preparation''~(2023).
\newblock  \href{http://arxiv.org/abs/2303.18224}{arXiv:2303.18224}.

\bibitem{QGS2}
Chi-Fang Chen, Michael~J Kastoryano, and Andr{\'a}s Gily{\'e}n.
\newblock ``An efficient and exact noncommutative quantum {G}ibbs
  sampler''~(2023).
\newblock  \href{http://arxiv.org/abs/2311.09207}{arXiv:2311.09207}.

\bibitem{QGS3}
Andr{\'a}s Gily{\'e}n, Chi-Fang Chen, Joao~F Doriguello, and Michael~J
  Kastoryano.
\newblock ``Quantum generalizations of {G}lauber and {M}etropolis
  dynamics''~(2024).
\newblock  \href{http://arxiv.org/abs/2405.20322}{arXiv:2405.20322}.

\bibitem{QGS4}
Zhiyan Ding, Bowen Li, and Lin Lin.
\newblock ``Efficient quantum {G}ibbs samplers with
  {K}ubo--{M}artin--{S}chwinger detailed balance condition''.
\newblock \href{https://dx.doi.org/10.1007/s00220-025-05235-3}{Communications
  in Mathematical Physics {\bf 406}, 67}~(2025).

\bibitem{sachdev1999quantum}
Subir Sachdev.
\newblock ``Quantum phase transitions''.
\newblock \href{https://dx.doi.org/10.1088/2058-7058/12/4/23}{Physics world
  {\bf 12}, 33}~(1999).

\bibitem{zeng2019quantum}
Bei Zeng, Xie Chen, Duan-Lu Zhou, Xiao-Gang Wen, et~al.
\newblock ``Quantum information meets quantum matter''.
\newblock \href{https://dx.doi.org/10.1007/978-1-4939-9084-9}{Springer}.
  ~(2019).

\bibitem{bardet2023rapid}
Ivan Bardet, {\'A}ngela Capel, Li~Gao, Angelo Lucia, David
  P{\'e}rez-Garc{\'\i}a, and Cambyse Rouz{\'e}.
\newblock ``Rapid thermalization of spin chain commuting {H}amiltonians''.
\newblock \href{https://dx.doi.org/10.1103/PhysRevLett.130.060401}{Physical
  Review Letters {\bf 130}, 060401}~(2023).

\bibitem{gardiner2004quantum}
Crispin Gardiner and Peter Zoller.
\newblock ``Quantum noise: a handbook of {M}arkovian and non-{M}arkovian
  quantum stochastic methods with applications to quantum optics''.
\newblock \href{https://dx.doi.org/10.1007/978-3-662-04103-1}{Springer Science
  \& Business Media}. ~(2004).

\bibitem{nathan2020universal}
Frederik Nathan and Mark~S Rudner.
\newblock ``Universal {L}indblad equation for open quantum systems''.
\newblock \href{https://dx.doi.org/10.1103/PhysRevB.102.115109}{Physical Review
  B {\bf 102}, 115109}~(2020).

\bibitem{berger2005}
Noam Berger, Claire Kenyon, Elchanan Mossel, and Yuval Peres.
\newblock ``Glauber dynamics on trees and hyperbolic graphs''.
\newblock \href{https://dx.doi.org/10.1007/s00440-004-0369-4}{Probability
  Theory and Related Fields {\bf 131}, 311--340}~(2005).

\bibitem{knill2000theory}
Emanuel Knill, Raymond Laflamme, and Lorenza Viola.
\newblock ``Theory of quantum error correction for general noise''.
\newblock \href{https://dx.doi.org/10.1103/PhysRevLett.84.2525}{Physical Review
  Letters {\bf 84}, 2525}~(2000).

\bibitem{zanardi2000stabilizing}
Paolo Zanardi.
\newblock ``Stabilizing quantum information''.
\newblock \href{https://dx.doi.org/10.1103/PhysRevA.63.012301}{Physical Review
  A {\bf 63}, 012301}~(2000).

\bibitem{kraus2008preparation}
Barbara Kraus, Hans~P B{\"u}chler, Sebastian Diehl, Adrian Kantian, Andrea
  Micheli, and Peter Zoller.
\newblock ``Preparation of entangled states by quantum {M}arkov processes''.
\newblock \href{https://dx.doi.org/10.1103/PhysRevA.78.042307}{Physical Review
  A {\bf 78}, 042307}~(2008).

\bibitem{lidar2003}
Daniel~A Lidar and K~Birgitta~Whaley.
\newblock ``Decoherence-free subspaces and subsystems''.
\newblock In Irreversible quantum dynamics.
\newblock \href{https://dx.doi.org/10.1007/3-540-44874-8_5}{Pages 83--120}.
\newblock Springer~(2003).

\bibitem{chesi2010thermodynamic}
Stefano Chesi, Daniel Loss, Sergey Bravyi, and Barbara~M Terhal.
\newblock ``Thermodynamic stability criteria for a quantum memory based on
  stabilizer and subsystem codes''.
\newblock \href{https://dx.doi.org/10.1088/1367-2630/12/2/025013}{New Journal
  of Physics {\bf 12}, 025013}~(2010).

\bibitem{johansson2012}
J~Robert Johansson, Paul~D Nation, and Franco Nori.
\newblock ``{Q}u{T}i{P}: An open-source {P}ython framework for the dynamics of
  open quantum systems''.
\newblock \href{https://dx.doi.org/10.1016/j.cpc.2012.02.021}{Computer physics
  communications {\bf 183}, 1760--1772}~(2012).

\bibitem{okamoto1992fluid}
Kiyomi Okamoto and Kiyohide Nomura.
\newblock ``Fluid-dimer critical point in {S} = 12 antiferromagnetic
  {H}eisenberg chain with next nearest neighbor interactions''.
\newblock \href{https://dx.doi.org/10.1016/0375-9601(92)90823-5}{Physics
  Letters A {\bf 169}, 433--437}~(1992).

\bibitem{caspers1982some}
WJ~Caspers and W~Magnus.
\newblock ``Some exact excited states in a linear antiferromagnetic spin
  system''.
\newblock \href{https://dx.doi.org/10.1016/0375-9601(82)90603-x}{Physics
  letters A {\bf 88}, 103--105}~(1982).

\bibitem{hamermesh2012}
M.~Hamermesh.
\newblock ``Group theory and its application to physical problems''.
\newblock \href{https://dx.doi.org/10.1063/1.3050758}{Dover Books on Physics}.
  Dover Publications. ~(2012).

\bibitem{bransden2003}
B.H. Bransden and C.J. Joachain.
\newblock ``Physics of atoms and molecules''.
\newblock Pearson Education. Prentice Hall. ~(2003).
\newblock  url:~\url{https://books.google.dk/books?id=i5IPWXDQlcIC}.

\bibitem{campos2021}
Jorge~A Campos-Gonzalez-Angulo, Raphael~F Ribeiro, and Joel Yuen-Zhou.
\newblock ``Generalization of the {T}avis--{C}ummings model for multi-level
  anharmonic systems''.
\newblock \href{https://dx.doi.org/10.1088/1367-2630/ac00d7}{New Journal of
  Physics {\bf 23}, 063081}~(2021).

\bibitem{arovas1989two}
Daniel~P Arovas.
\newblock ``Two exact excited states for the {S} = 1 {AKLT} chain''.
\newblock \href{https://dx.doi.org/10.1016/0375-9601(89)90921-3}{Physics
  Letters A {\bf 137}, 431--433}~(1989).

\bibitem{faddeev1984spectrum}
Lyudvig~Dmitrievich Faddeev and Leon~Armenovich Takhtadzhyan.
\newblock ``Spectrum and scattering of excitations in the one-dimensional
  isotropic {H}eisenberg model''.
\newblock \href{https://dx.doi.org/10.1007/BF01087245}{Journal of Soviet
  Mathematics {\bf 24}, 241--267}~(1984).

\bibitem{doikou1999parity}
Anastasia Doikou and Rafael~I Nepomechie.
\newblock ``Parity and charge conjugation symmetries and {S} matrix of the
  {XXZ} chain''.
\newblock Statistical Physics on the Eve of the Twenty-First Century, M.
  Batchelor and L. Wille, edsPages 391--411~(1999).
\newblock  \href{http://arxiv.org/abs/hep-th/9810034}{arXiv:hep-th/9810034}.

\bibitem{wen2003quantum}
Xiao-Gang Wen.
\newblock ``Quantum orders in an exact soluble model''.
\newblock \href{https://dx.doi.org/10.1103/PhysRevLett.90.016803}{Physical
  review letters {\bf 90}, 016803}~(2003).

\bibitem{matthies2022programmable}
Anne Matthies, Mark Rudner, Achim Rosch, and Erez Berg.
\newblock ``Programmable adiabatic demagnetization for systems with trivial and
  topological excitations''.
\newblock \href{https://dx.doi.org/10.22331/q-2024-10-23-1505}{Quantum {\bf 8},
  1505}~(2024).

\bibitem{kishony2023gauged}
Gilad Kishony, Mark~S Rudner, Achim Rosch, and Erez Berg.
\newblock ``Gauged cooling of topological excitations and emergent fermions on
  quantum simulators''.
\newblock \href{https://dx.doi.org/10.1103/PhysRevLett.134.086503}{Physical
  Review Letters {\bf 134}, 086503}~(2025).

\bibitem{andersen2024thermalization}
Trond~I Andersen, Nikita Astrakhantsev, Amir~H Karamlou, Julia Berndtsson,
  Johannes Motruk, Aaron Szasz, Jonathan~A Gross, Alexander Schuckert, Tom
  Westerhout, Yaxing Zhang, et~al.
\newblock ``Thermalization and criticality on an analogue--digital quantum
  simulator''.
\newblock \href{https://dx.doi.org/10.1038/s41586-024-08460-3}{Nature {\bf
  638}, 79--85}~(2025).

\bibitem{lloyd2024quasiparticle}
Jerome Lloyd, Alexios~A Michailidis, Xiao Mi, Vadim Smelyanskiy, and Dmitry~A
  Abanin.
\newblock ``Quasiparticle cooling algorithms for quantum many-body state
  preparation''.
\newblock \href{https://dx.doi.org/10.1103/PRXQuantum.6.010361}{PRX Quantum
  {\bf 6}, 010361}~(2025).

\bibitem{nakamura2002order}
Masaaki Nakamura and Synge Todo.
\newblock ``Order parameter to characterize valence-bond-solid states in
  quantum spin chains''.
\newblock \href{https://dx.doi.org/10.1103/PhysRevLett.89.077204}{Physical
  review letters {\bf 89}, 077204}~(2002).

\bibitem{wang2020classes}
Dong-Sheng Wang.
\newblock ``Classes of topological qubits from low-dimensional quantum spin
  systems''.
\newblock \href{https://dx.doi.org/10.1016/j.aop.2019.168015}{Annals of Physics
  {\bf 412}, 168015}~(2020).

\end{thebibliography}

\onecolumn
\appendix

%In the appendix, we (i) look more carefully at the contribution from the Hamiltonian term $Q$, and (ii) discuss how the metastaibility behaves as the system size grows. 

\section{Contribution from $Q$}
\label{app:Q}

We derive the bound on $||\kappa |\psi\rangle||$. To start with, we break up  the norm as 
\begin{equation}
    ||\kappa |\psi\rangle||\leq \sum_a ||Q_a |\psi\rangle||+\frac{1}{2}||L^\dag_a L_a|\psi\rangle||,\label{eqn:tobebounded}
\end{equation}
where we have defined 
\begin{equation}
    Q_a=\frac{i}{2}\sum_{ij}\tanh\left(\frac{\beta(e_i-e_j)}{4}\right)\langle e_i|L_a^\dag L_a|e_j\rangle |e_i\rangle\langle e_j|.
\end{equation}

Now lets consider the blind qubit subspace ${\rm span}\{|e_0\rangle,|e_l\rangle\}$, for some eigenstate $|e_l\rangle$ of $H$; i.e. $\langle e_l|S_a|e_j\rangle=0$ for all Paulis $S_a$ and all $j\leq l$.  We showed in the main text,  
\begin{equation}
    ||L_a|e_l\rangle||\leq O(e^{-\beta \Delta/2}).
\end{equation}
The second term in Eqn. (\ref{eqn:tobebounded}) can therefore be bounded simply as 
\begin{equation}
    \frac{1}{2}||L^\dag_a L_a|e_l\rangle||\leq \frac{1}{2}||L^\dag_a||~ || L_a|e_l\rangle||\leq O(e^{-\beta \Delta/2}),
\end{equation}
since $||L^\dag_a||=O(1)$ \cite{QGS4}. The first term can similarly be bounded by noting that 
\begin{eqnarray}
    |\langle e_j |L_a^\dag L_a |e_l\rangle| \leq O(e^{-\beta \Delta/2}),
\end{eqnarray}
for the same reasons as above, and thus
\begin{eqnarray}
    ||\change{Q_a}|e_l\rangle|| &\leq&\frac{1}{2}\sum_j \tanh\left(\frac{\beta(e_j-e_l)}{4}\right)|\langle e_j |L_a^\dag L_a |e_l\rangle|\\
    &\leq&\frac{1}{2}\sum_j |\langle e_j |L_a^\dag L_a |e_l\rangle|\leq O(e^{-\beta \Delta/2}).
\end{eqnarray}
By summing up all Pauli operators, we get a scaling $||\kappa |e_l\rangle||= O(N e^{-\beta \Delta/2})$.

\section{ The large system limit}

In the main text, we showed that if a system has a blind subspace, then this subspace will be metastable for a time scaling as $\Omega(N^{-1} e^{\beta \Delta/2})$. Our arguments relied on numerical evidence on small systems ($N<16$). Below, we explore to what extend some of our conclusions might break down for larger system sizes.

\subsection{System size dependence of $\Delta$.}
Leakage out of the blind subspace is bounded by the quantity $e^{-\beta \Delta/2}$, where $\Delta=\min\{e_{j+1}-e_j,e_p -e_0\}$ is the smallest energy gap from a state in the blind subspace to a higher-energy state outside of it. To ensure slow leakage out of the subspace, we must therefore require $\beta \Delta \gg 1$. Thus, the size of the gap is important for practical applications, since $\Delta^{-1}$ sets a lower bound for the required inverse temperature $\beta$, as well as determines the strength of the suppression at low temperatures. To investigate how this quantity evolves as the system size grows, Lanczos-based numerical diagonalization was used to determine the low-energy spectrum and calculate the gap for system-sizes up to $N=20$.  
\begin{figure}[ht]
    \centering
    \includegraphics[width=0.6\linewidth]{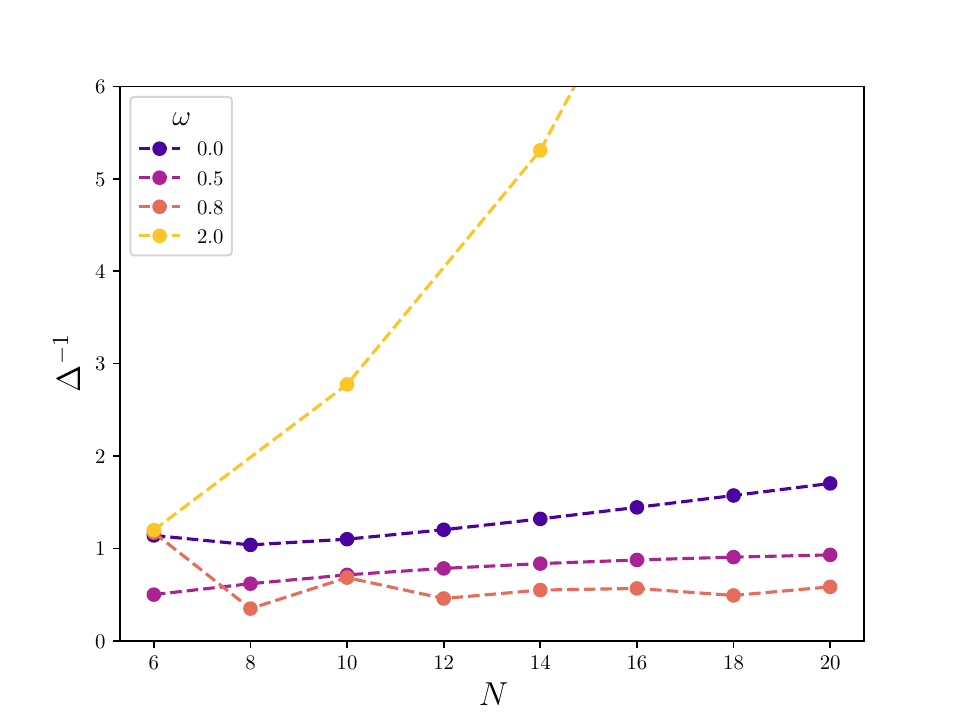}
    \caption{Magnitude of the gap above the blind subspace ($\Delta$) as a function of the system size $N$ for different values of $\omega$. Dotted lines have been added between data points to guide the eyes.}
    \label{fig:gap_sizes}
\end{figure}
In Fig. \ref{fig:gap_sizes}, the inverse gap above the blind subspace is plotted for representative points in the qubit VBS subspace ($\omega = 0.8$), in the 'Heisenberg' subspace ($\omega = 0$), and in the qutrit VBS subspace ($\omega = 2$). The gap for the  Majumdar-Gosh point ($\omega = 0.5$) is also plotted.  As can be seen from this figure, the gaps for  values of $\omega$ in the 'Heisenberg' and qubit VBS regions are relatively constant, with some indications of slowly decreasing gaps as system size is scaled up. However, in the qutrit-region $\omega\geq 1.5$,  the gap closes rapidly, corresponding to the disappearance of the protected subspace.  Note that the qutrit blind subspace is only observed for $N  \equiv 2\, (\mathrm{mod}\, 4)$, meaning the observed vanishing gap is consistent with an expectation that $N  \equiv 2\, (\mathrm{mod}\, 4)$ and $N  \equiv 0\, (\mathrm{mod}\, 4)$ should behave equivalently in the thermodynamic limit.

\subsection{Contributions from the bulk of the spectrum}
\label{app:boltzmann_sum_alt}
In the main text, it was conjectured that for low enough temperatures, the following holds as the system size is increased:
\begin{align}
    \change{\mathcal{K}_a}(N, \beta, \omega) := \sum_{k>j+1}e^{-\beta(e_k-e_{j+1})/2}|\langle e_k|S_a|e_j\rangle| \sim O(1) \; ,
    \label{eq:K_reminder}
\end{align}
where $\{e_k\}$ are the eigenenergies of the system numbered in increasing order, and $j$ is the index of the high-energy state of the blind qubit subspace. Here, we provide numerical evidence to support this claim, and explore what constitutes sufficiently low temperatures.

\begin{figure}[ht]
    \centering
    \includegraphics[width=0.5\linewidth]{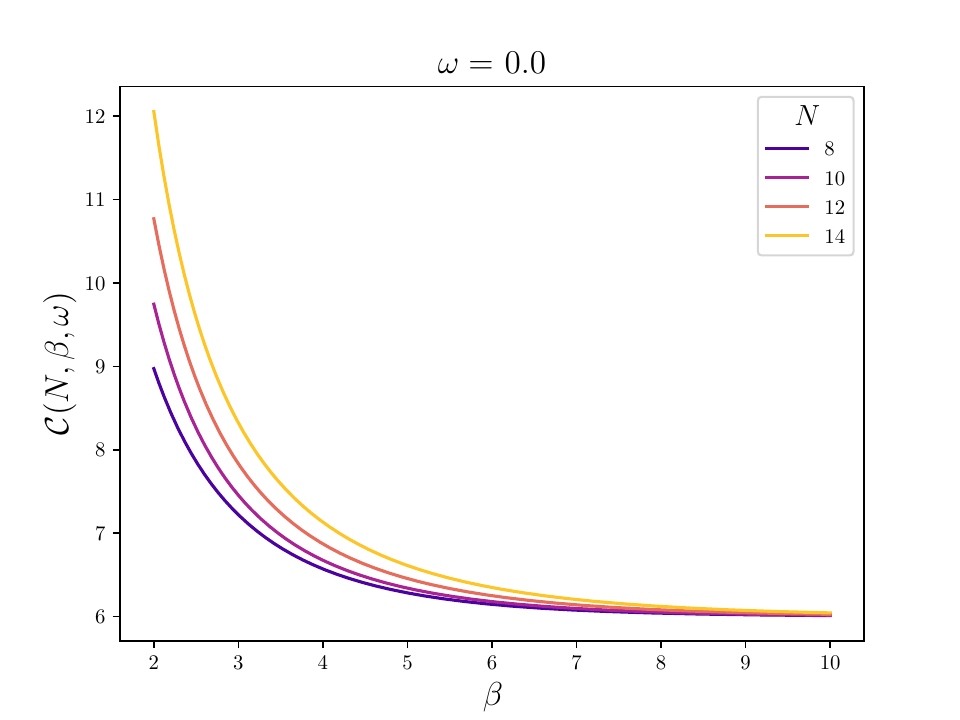}
    \caption{Evaluation of $\mathcal{C}(N, \beta, \omega)$ as a function of $\beta$ for $\omega=0$ and various system sizes. }
    \label{fig:decay_of_C}
\end{figure}

First, note that 
\begin{align}
    \change{\mathcal{K}_a}(N, \beta, \omega)\leq \mathcal{C}(N, \beta, \omega) := \sum_{k>j+1}e^{-\beta(e_k-e_{j+1})/2}  \; .
    \label{eq:K-C_Bound}
\end{align}
If the degeneracy for the energy level that the $(j+1)$th state belongs to is $D(N,\omega)$, the sum takes the form
\begin{align}
    \mathcal{C}(N, \beta, \omega) = D(N, \omega) + \sum_{k>(j+D)}e^{-\beta(e_k-e_{j+1})/2}
\end{align}
where now the energies in the exponent of the sum are all strictly positive by definition. Thus, in the limit of $\beta \rightarrow \infty$, $\mathcal{C}(N, \beta, \omega)$ and $D(N, \omega)$ coincide. Arguing that $\mathcal{C} \sim O(1)$ therefore boils down to arguing that a) $D(N, \omega)\sim O(1)$, i.e., that the degeneracies of the low-energy spectrum does not grow with system size, and b) the temperature scale where $\mathcal{C} \simeq D$ remains constant, or at the very least only shrinks at a relatively slow rate. To get an intuition for both of these questions, the evolution of $\mathcal{C}$ towards $D$ as $\beta$ is increased is depicted in Fig. \ref{fig:decay_of_C} for $\omega=0$, i.e., in the Heisenberg phase. 
As can be seen from this figure, the limit of $\mathcal{C}$ as $\beta$ is increased is unchanged as $N$ is increased, indicating that the degeneracy $D$ does not depend on the system size for even-size systems. This was additionally numerically verified up to a system size of $N=22$ using the Lanczos algorithm for finding the low-energy spectrum. A similar independence of $D$ to system size was observed across a range of $\omega$ in the relevant regimes, including in the qutrit-regime $\omega \gtrsim 1.5$ for systems of the size where the protected qutrit subspace is present, i.e., systems where $N  \equiv 2\, (\mathrm{mod}\, 4)$.

The results above provide support to the first of the two claims, namely that the degeneracies of the low-energy spectrum is independent of system size, meaning the  low-temperature limit of $\mathcal{C}(N, \beta, \omega)$ is also independent of system size. However, as can be seen in Fig. \ref{fig:decay_of_C}, the rate at which the asymptotic value is reached does depend on the system size, with larger systems typically displaying a slower decay. One way to quantify this slowdown is to investigate how large a value of $\beta$ is required to achieve a certain value of $\mathcal{C}$,
\begin{align}
    \beta_{\mathcal{C}}^*(N, \omega, K) = \inf \{ \beta : \mathcal{C}(N, \beta, \omega) \leq K \} \; .
\end{align}
\begin{figure}[ht]
    \centering
    \includegraphics[width=0.95\linewidth]{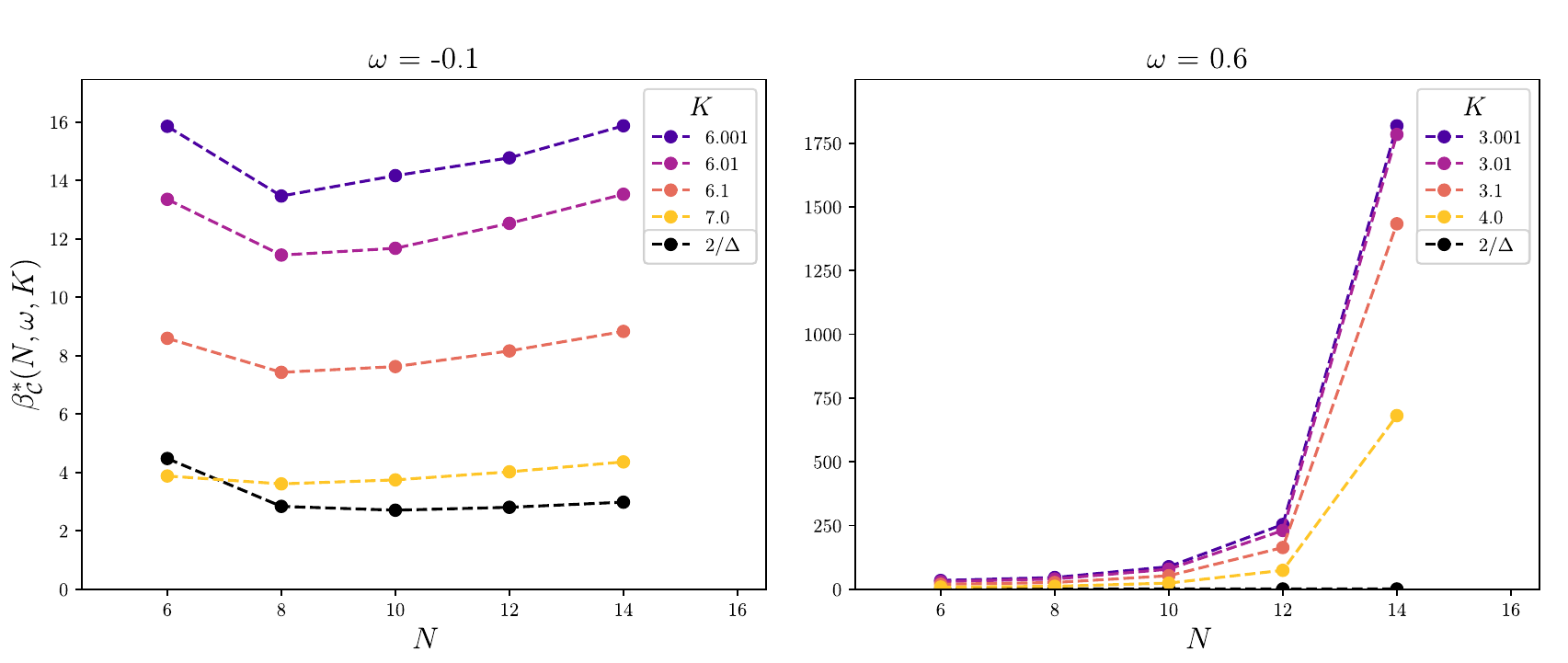}
    \caption{Numerical calculations of $\beta_{\mathcal{C}}^*(N, \omega, K)$, the value of $\beta$ required to achieve a given value $K$ of the upper bound $\mathcal{C}$ from Eqn. (\ref{eq:K-C_Bound}). The corresponding values for the inverse gap have been added for reference, and dotted lines added to guide the eyes.}
    \label{fig:required_beta_C}
\end{figure}
However, numerics indicate that the bound in Eqn.~\eqref{eq:K-C_Bound} is loose in some regimes. Specifically, picking values of $\omega$ that lie deep in the two meta-stable phases, while avoiding the special points ($\omega=0$ and $\omega=0.5$), yields the results in Fig.~\ref{fig:required_beta_C} for $\beta_{\mathcal{C}}^*$. This highlights that achieving $\mathcal{C} \simeq D$ requires rapidly increasing $\beta$ with system size in some regimes of $\omega$. On the other hand, defining a similar quantity for the full sum to be bounded,
\begin{align}
    \beta_{\mathcal{K}}^*(N, \omega, K) &= \inf \{ \beta : \mathcal{K}(N, \beta, \omega)  \leq K \} \\
   \change{ \mathcal{K}(N, \beta, \omega)} &\change{= \sum_{S_a \in \{ X_0, Y_0, Z_0\}} \mathcal{K}_a(N, \beta, \omega)  \; ,}
\end{align}
\begin{figure}[ht]
    \centering
    \includegraphics[width=0.95\linewidth]{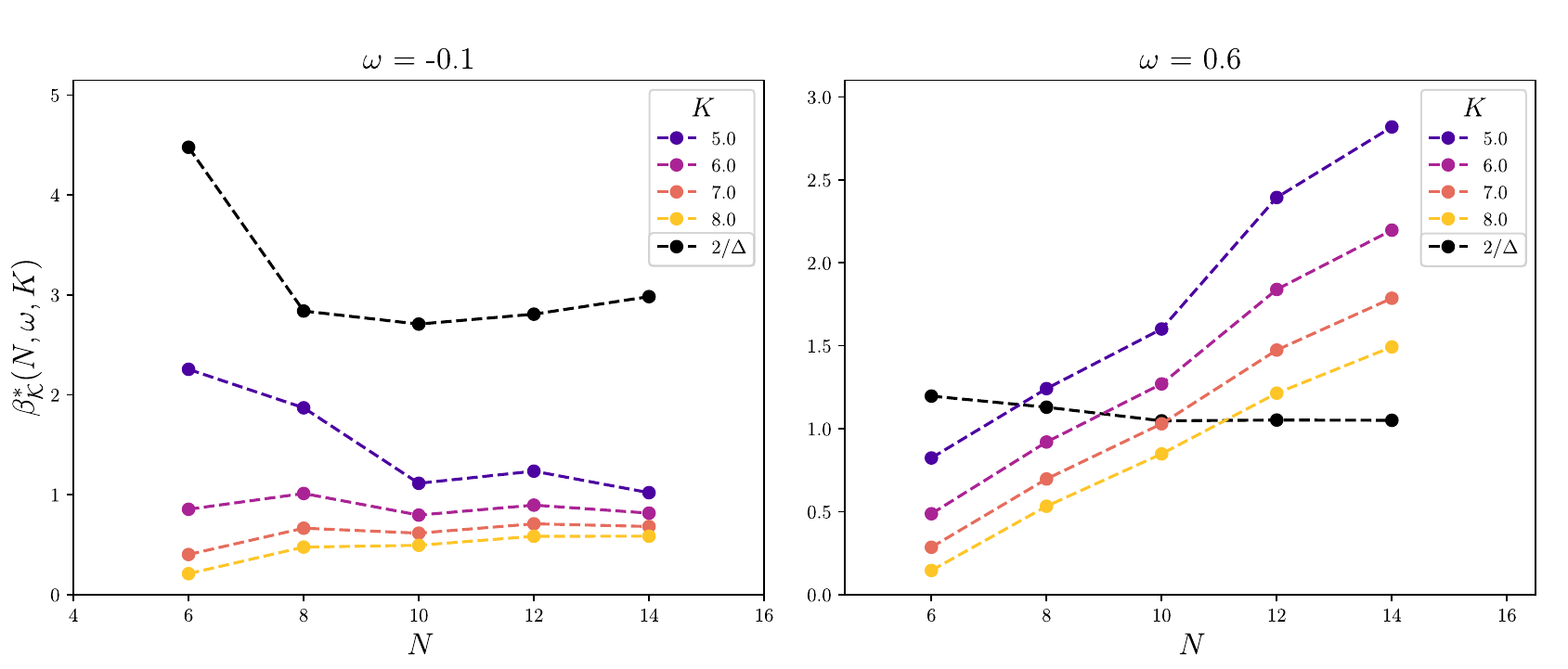}
    \caption{Numerical calculations of $\beta_{\mathcal{K}}^*(N, \omega, K)$, the value of $\beta$ required to achieve a given value $K$ of the quantity $\mathcal{K}$ in Eqn. (\ref{eq:K_reminder}). The corresponding values for the inverse gap have been added for reference, and dotted lines added to guide the eyes.}
    \label{fig:required_beta_K}
\end{figure}
yields the results in Fig.~\ref{fig:required_beta_K}, showing a slower increase as the system size is increased. \change{Note that for simplicity, the effect of $X$, $Y$ and $Z$-type operators has been combined, and translational invariance implicitly used to allow a focus on operators on the first qubit.} The exact way that $\beta_{\mathcal{K}}^*$ evolves as $N$ grows depends on $\omega$ and $K$, but generally follows similar patterns of approximately linear growth with $N$ for the investigated system sizes. This supports the claim that $\mathcal{K}$ can be kept constant as the system size is increased using at most a linear increase in $\beta$, also beyond the regimes where it follows from a bound on the simpler quantity $\mathcal{C}$.

Beyond supporting the scaling behaviour, similar numerics can be used to investigate the full value of the sum in Eqn.~\eqref{eq:jump_op_norm}. Specifically, we can define the inverse temperature needed to reduce the combined operator-norm below a certain threshold,
\begin{align}
    \beta^*(N, \omega, K) = \inf \{ \beta : 2 \pi N e^{-\beta \Delta/2}\mathcal{K}(N, \beta, \omega) \leq K \} \; ,
\end{align}
\begin{figure}[htb]
    \centering
    \includegraphics[width=0.99\linewidth]{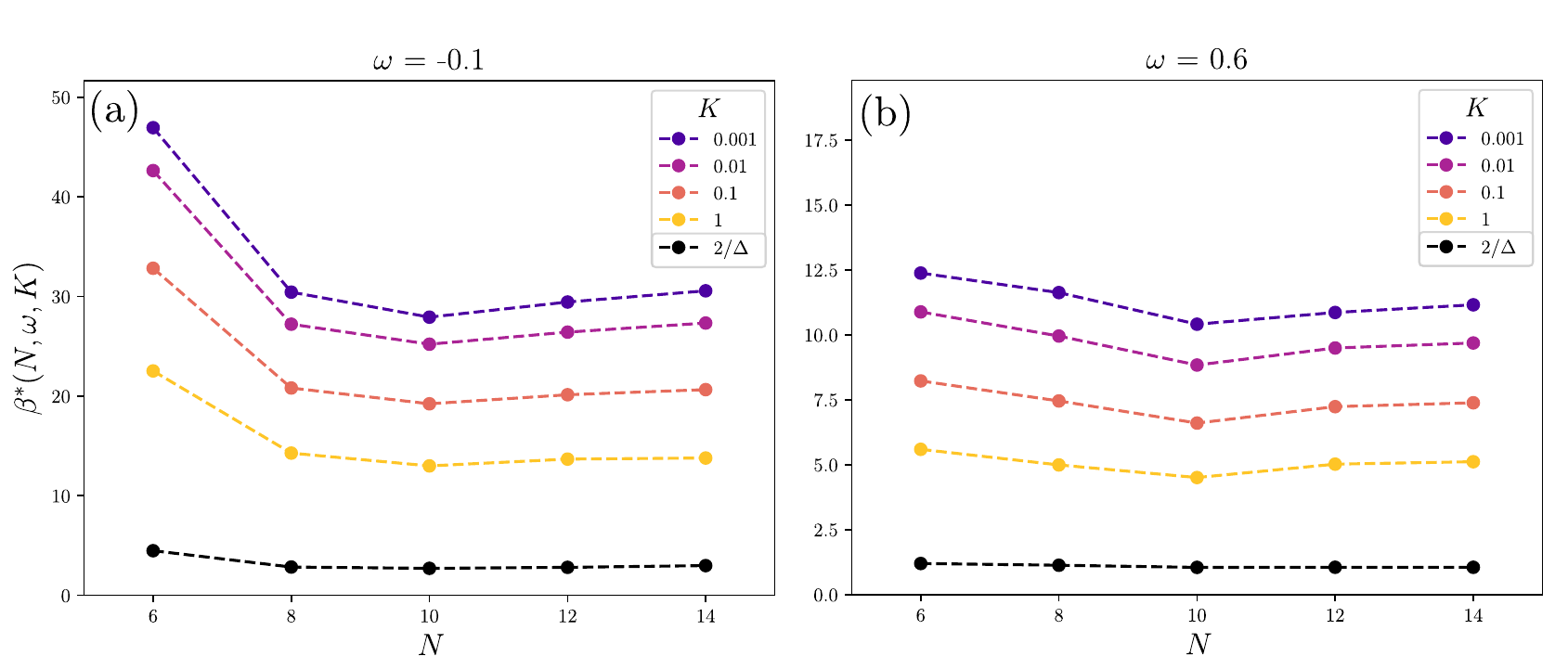}
    \caption{Numerical calculations of $\beta^*(N, \omega, K)$, the value of $\beta$ required to achieve a given suppression $K$ in Eqn. (\ref{eq:jump_op_norm}). The corresponding values for the inverse gap have been added for reference, and dotted lines added to guide the eyes.}
    \label{fig:required_beta}
\end{figure}
which indicates more directly the temperature required to reach a certain level of protection of the blind subspace. This quantity is depicted on Fig. \ref{fig:required_beta} for $\omega=-0.1$ and $\omega=0.6$. While it is difficult to make any strong claims based on the figure, the value of $\beta$ necessary to reach a given protection does appear to increase slowly beyond $N\approx 10$. A similar behavior is seen for other choices of $\omega$. The exact way that $\beta^*$ evolves as $N$ grows depends on $\omega$ and $K$, but often follows a pattern where stricter (lower) values of $K$ have a positive curvature, while more lenient (higher) values of $K$ have a negative curvature. One exception is near $\omega \simeq 0.8$, where oscillatory finite-size effects are observed, similarly to the behaviour of the gaps in Fig. \ref{fig:gap_sizes}.

In conclusion, numerics suggest that if $\beta \geq \beta^*$  for a $\beta^*$ that increases sub-linearly with system size, the scaling $\mathcal{K} = O(1)$ is achieved, and all of the decoherence-inducing contributions are therefore bounded by $O(e^{-\beta \Delta/2})$. Since the inverse gap also scales approximately linearly (see Fig. \ref{fig:gap_sizes}), both of these effects enforce an approximately linearly increasing $\beta$ with system size.

\subsection{Phase transition points as the system scales up}
\begin{figure}[ht]
    \centering
    \includegraphics[width=.5\linewidth]{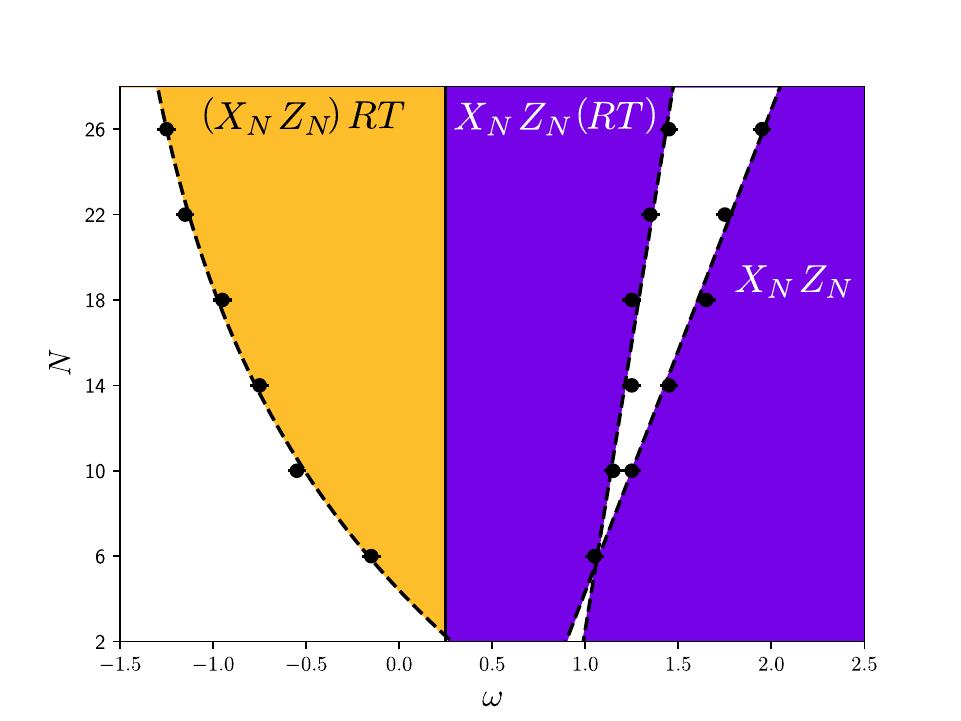}
    \caption{Location of the transitions between metastable phases.  From left to right, the second region hosts a protected qubit, while the fourth hosts a protected qutrit. Fits are either linear or exponential in the system size, depending on the observed behaviour. For the exponential fit, a limit of $\omega^*_L \simeq -1.647$ is predicted, with the characteristic scale of $k^{-1} \simeq 15$ in the exponential decay $\omega(N) = \omega^*_L + (\omega_L^{(0)} - \omega^*_L) e^{-kN}$.}
    \label{fig:phase_transition_points}
\end{figure}
A compelling feature of the results presented here is that the transitions  into and out of metastable subspaces do not appear to coincide with the  ground state phase diagram for finite $N$. To test whether this feature survives as the chain length grows, we plot the (approximate) locations of the transition  points in Fig. \ref{fig:phase_transition_points}. The numerics suggest that the locations of transitions to the qubit and qutrit metastable subspaces grow linearly with $N$, strongly suggesting that the qubit metastable subspace will cover the whole phase  in the thermodynamic limit. The 'Heisenberg' subspace is also moving left as the system size increases, but it is difficult to know whether it will continue expanding as the system size grows further. 

If the two phases extend out to infinity, this would suggest that the metastability 'phase diagram' coincides with the ground state phase diagram for the $J_-J_2$ model. In parallel experiments, that are not reported here, we found that this is clearly not the case for the XXZ Heisenberg model, which undergoes a metastable transition at zero ZZ field strength, which does not coincide with the ground state phase transition of the model. 

Finally, in view of the observation above, that the critical temperature $\beta^*$ above which there is exponential survival of the blind subspace grows with system size, the effects that we report are most pronounced for finite chains.

\section{Sensitivity to symmetry-breaking perturbations}
\label{app:Sensitivity}
As outlined in the main text, the blind subspaces are formed due to protection caused by the symmetries of the system. Specifically, the symmetries $R$ and $T$ protect the fourth excited state of the $\omega < \omega^*$ Heisenberg-like phase, while the first excited state of the $\omega^* < \omega \lesssim 1$ VBS-like phase is independently and redundantly protected by both the symmetries $Z_N$ and $X_N$ and the symmetries $R$ and $T$ (See, e.g., Fig. \ref{fig:two} of the main text). \\

As mentioned in the main text, this protection is expected to break down if perturbations breaking the symmetries are added to the system Hamiltonian. Below, we will elucidate this effect for a small selection of local perturbations, and provide numerical results on the performance of the protection when the system is subjected to disordered local fields. The central figure of merit will be the overlaps induced by the symmetry breaking:
\begin{align}
    \tau_n^{A} = \sum_{i<j} \left| \left< e_i \right| P_n \left| e_j \right> \right|^2  & & P\in \{X, Y, Z\}, \; n \in \{1, 2 \dots N\} \; .
\end{align}
In the case of no perturbations, the symmetries enforce these matrix elements to be zero. On the other hand, non-zero values implies non-zero contributions to the sum in Eq. \eqref{eq:jump_op_norm} in the main text, and thus a breakdown of metastability. The results below are based on analyzing these quantities numerically for system sizes of $N=8$ to $N=14$ at $\omega = -0.1$ (Heinsenberg-like phase) and $\omega = 0.6$ (VBS-like phase), though the conclusions are not expected to depend sensitively on these parameters. 

\subsection{Global homogeneous field}
Consider first a homogeneous local field applied identically to every site. Due to the symmetry of the un-perturbed model, we assume without loss of generality that the field points in the $Z$ direction, i.e., we consider the perturbation
\begin{align*}
    H_p(x) = x \sum_{n = 1}^N Z_n \; .
\end{align*}
This perturbation breaks the $X_N$ symmetry, but leaves the rest of the symmetries intact. This means that the $R$ and $T$-based protection of both the Heisenberg phase and VBS phase remains intact, with the relevant blind subspaces remaining blind and fully protected. The breakdown of the protection thus only occurs once the perturbation becomes sufficiently strong to induce (avoided) crossings in the spectrum of the Hamiltonian, which for $N=12$ occurs at $x \sim 0.4$ for the Heisenberg phase and $x \sim 0.9$ for the VBS phase.

\subsection{Local single-site perturbation}
\begin{figure}[htbp]
\centering
    \begin{subfigure}[b]{0.49\textwidth}
    \centering
    \includegraphics[scale=0.56]{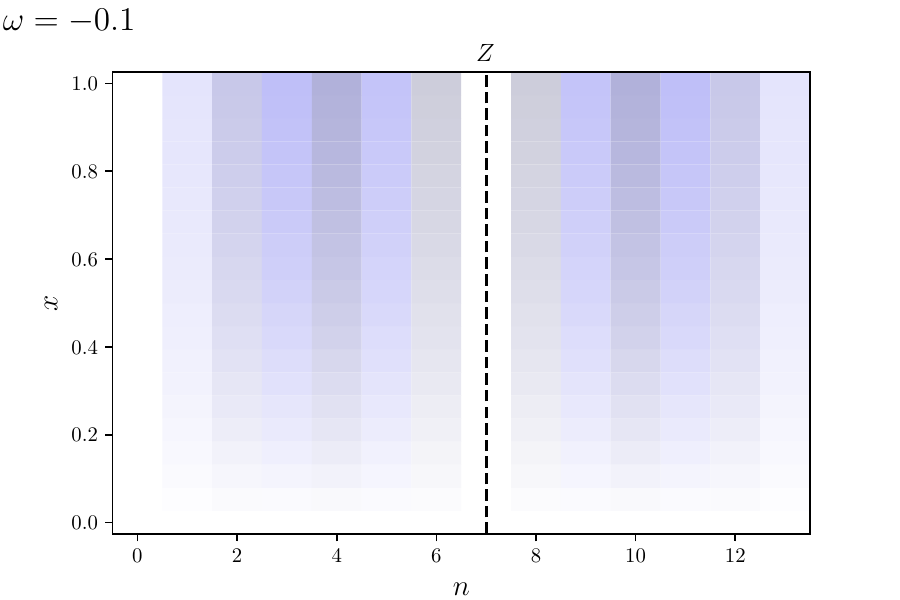}
    \end{subfigure}
    ~
    \begin{subfigure}[b]{0.49\textwidth}
    \centering
    \includegraphics[scale=0.56]{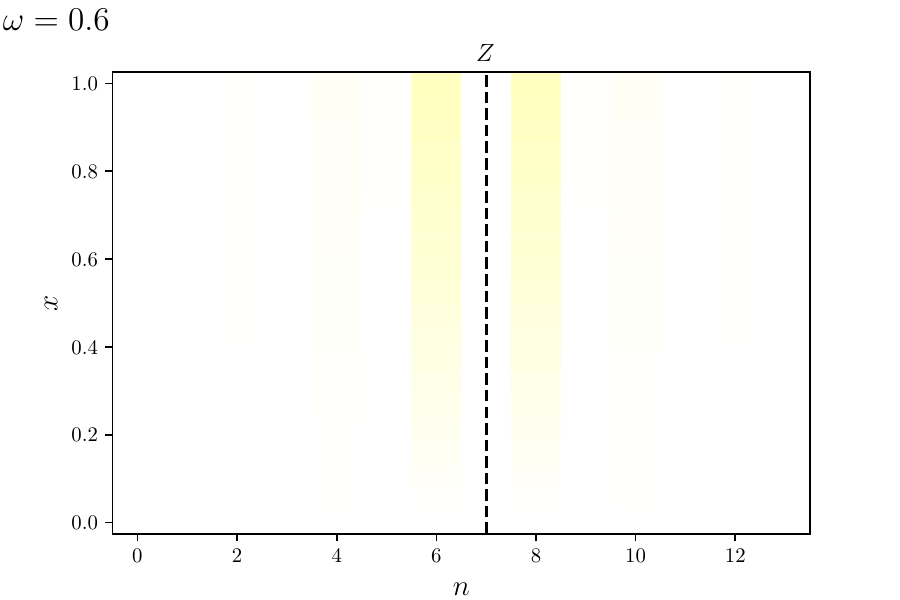}
    \end{subfigure}
    \caption{Sensitivities caused by a Hamiltonian perturbation $x \, Z_{7}$ on a chain of length $N=14$ in both the Heisenberg-like phase (left) and VBS-like phase (right). The $x$-axis labels the 14 qubits, and the $y$-axis represents the strength of the perturbation. The color of the patch at location $(n,x)$ is given by $(R, G, B)_{n,x} = \left(1 - \sqrt{\tau_n^X(x)}, 1 - \sqrt{\tau_n^Y(x)}, 1 - \sqrt{\tau_n^Z(x)}\right)$, i.e.,  each of the primary colors indicate robustness with respect to one of the three Paulis. Thus, white corresponds to full protection and vanishing sensitivities, $\tau_n^A = 0$, while darker colors indicate larger sensitivities.}
    \label{fig:Z_sensitivities}
\end{figure}

Consider now a perturbation of the form:
\begin{align}
    H_p(x) = x \, Z_{N/2} \; .
\end{align}
This perturbation breaks both the $X_N$ symmetry, the translation symmetry $T$, and most of the reflection symmetries. However, it does preserve symmetry under reflection about the axis through qubits $0$ and $N/2$. Denote this symmetry by $R'_0$. Using arguments similar to those of the main text, it is then straightforward to show that the matrix elements $\left<e_i \right| P_0  \left| e_j \right>$ and $\left<e_i \right| P_{N/2}  \left| e_j \right>$ related to single-qubit Paulis at the two qubits preserved by the reflection $R'_0$ can only be non-zero if $\lambda^{R'_0}_i = \lambda^{R'_0}_j$. Thus, $R'_0$ can provide blindness with respect to Paulis at those two qubits whenever $\lambda^{R'_0}_i \neq \lambda^{R'_0}_j$. In fact, looking at the $N=12$ symmetries of the low-energy states, the states $\left| e_4 \right>$ (Heisenberg) and $\left| e_0 \right>$ (VBS) indeed do have the requisite properties for this protection (See Table \ref{tab:example_app} below). In addition, protection against some local Paulis is implied by the remaining $Z_N$ symmetry by similar arguments. A depiction of the sensitivities (quantified by $\tau_n^A$) as the strength of the perturbation is scaled up is shown on Fig. \ref{fig:Z_sensitivities}, showing clearly the special protection of qubits 0 and $N/2$ implied by the $R'_0$ symmetry. Note also that the preserved $Z_N$ symmetry causes intact protection against $X$ and $Y$-noise in the VBS phase, thus causing the observed yellow tint in the VBS-phase from the combination of preserved $X$ (red) and $Y$ (green) robustness.

\begin{table}[ht]
\centering
\begin{tabular}{c | c | c || c | c | c }
$\omega=-0.1$ & $\lambda^{Z_N}$ & $\lambda^{R'_0}$ & $\omega=0.6$ & $\lambda^{Z_N}$ & $\lambda^{R'_0}$  \\
\hline\hline
$|e_0\rangle$ & $1$ & $1$ & $|e_0\rangle$ & $1$ & $1$ \\
\hline
$|e_1\rangle$ & $-1$ & $1$ & $|e_1\rangle$ & $1$ & $-1$   \\
\hline
$|e_2\rangle$ & $1$ & $1$  \\
\hline
$|e_3\rangle$ & $-1$ & $1$  \\
\hline
$|e_4\rangle$ & $1$ & $-1$  \\

\end{tabular}
\caption{Symmetry properties of the low-energy states of a system of size $N=12$ for both the Heisenberg-like phase (left) and the VBS-like phase (right).}
\label{tab:example_app}
\end{table}

\subsection{Two-site perturbation}
\begin{figure}[htbp]
\centering
    \centering
    \includegraphics[scale=0.56]{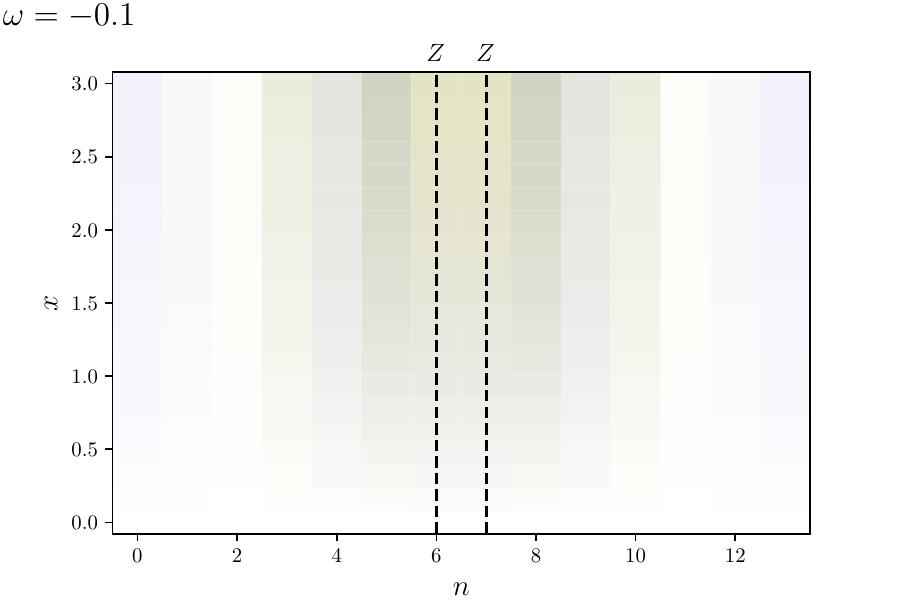}
    \caption{Sensitivities caused by a Hamiltonian perturbation $x \, Z_{6} \, Z_{7} $ on a chain of length $N=14$ in the Heisenberg-like phase. The $x$-axis labels the 14 qubits, and the $y$-axis represents the strength of the perturbation. The color of the patch at location $(n,x)$ is given by $(R, G, B)_{n,x} = \left(1 - \sqrt{\tau_n^X(x)}, 1 - \sqrt{\tau_n^Y(x)}, 1 - \sqrt{\tau_n^Z(x)}\right)$, i.e.,  each of the primary colors indicate robustness with respect to one of the three Paulis. Note the increased scale of the perturbation compared to Fig. \ref{fig:Z_sensitivities}.}
    \label{fig:ZZ_sensitivities}
\end{figure}

Consider next a perturbation of the form:
\begin{align}
    H_p(x) = x \, Z_{N/2-1} Z_{N/2} \; .
\end{align}
This perturbation breaks only the $T$ and $R$ symmetries, while preserving $Z_N$ and $X_N$. As a result, the blind subspace in the VBS-like phase is preserved under this perturbation. On the other hand, the Heisenberg-like phase relies on $T$ and $R$ for the protection of the blind subspace, and thus no longer exhibits protection. The induced sensitivities in the Heisenberg phase is depicted on Fig. \ref{fig:ZZ_sensitivities}. A few interesting observations from this figure is that the induced sensitivity is in general weaker than the one induced by the single-site perturbation, and that the induced sensitivity is predominantly localized around the site of the perturbation.

\subsection{Inhomogeneous field}
\begin{figure}[htbp]
\centering
    \begin{subfigure}[b]{0.49\textwidth}
    \centering
    \includegraphics[scale=0.56]{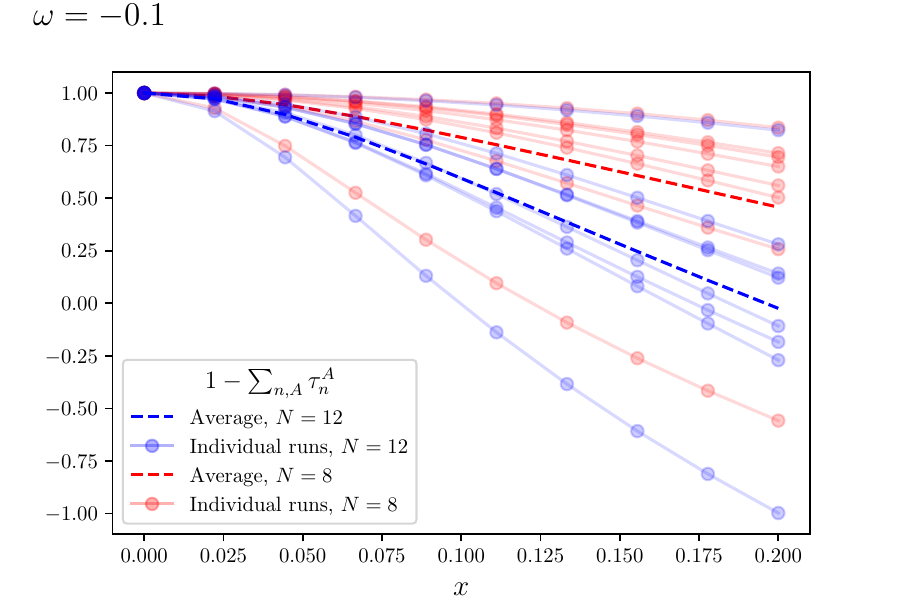}
    \end{subfigure}
    ~
    \begin{subfigure}[b]{0.49\textwidth}
    \centering
    \includegraphics[scale=0.56]{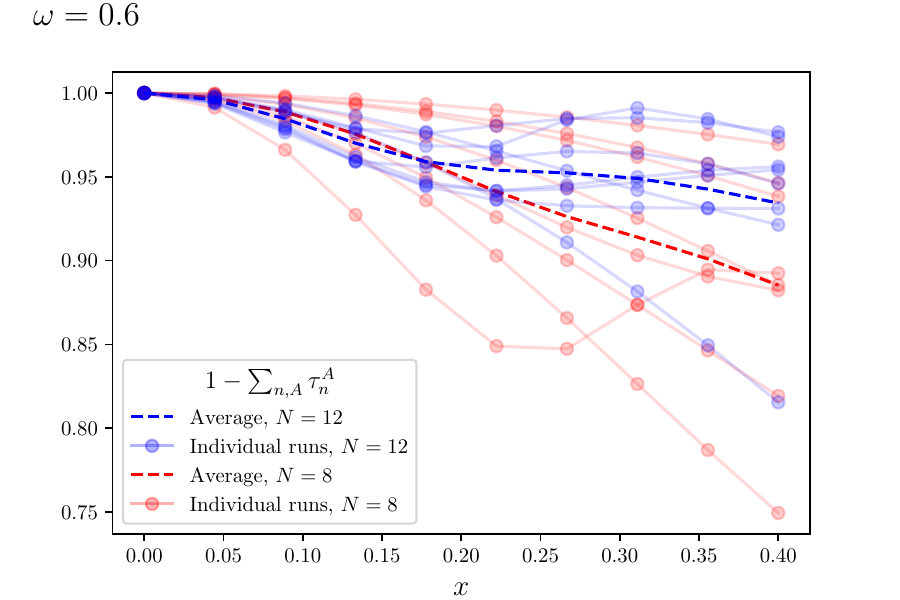}
    \end{subfigure}
    \caption{Robustness-proxy $1 - \sum_{n,A} \tau_n^A$ for different realizations of the noise disorder as the overall magnitude of the noise is increased. Identical noise realizations was used across both system sizes and in both phases to facilitate easier comparison. Note the different axes for the two plots.}
    \label{fig:disordered_sensitivities}
\end{figure}

The preceding examples provide evidence of the effects of specific perturbations breaking specific symmetries. A different but equally important question is the effect of less structured perturbations. Indeed, small random deviations of the implemented Hamiltonian from the fully symmetric theoretical model is likely to occur in any physical experiment investigating the observed effects. As a small preliminary investigation of the effect of such disordered perturbations on the observed protection, we consider now perturbations of the form:
\begin{align}
   H_p(x) = x \, \sum_{n} s_n \, Z_n \; ,
\end{align}
with each $s_n \in \mathbb{R}$ drawn independently from a normal distribution of mean zero and variance one. Tracking the evolution of the robustness proxy $1 - \sum_{n,A} \tau_n^A (x)$ as the overall magnitude $x$ of this in-homogeneous field is scaled up provides some evidence of the degree of robustness of the protection for different phases and system sizes. Numerical results from $8$ disorder realizations is shown on Fig. \ref{fig:disordered_sensitivities}. The main qualitative observations is that the VBS-like phase exhibits larger general robustness of the blind subspace (note the different axes), and that the Heisenberg-like phase seems to exhibit less robustness as system size increases.\\

An additional qualitative observation worth making is the presence of oscillatory revivals of the robustness for the VBS-like phase. By the definition of the robustness proxy, this indicates an oscillation in the magnitude of the matrix elements between the two states $\left| e_0 \right>$ and $\left| e_1 \right>$ for some noise realizations. In fact, similar oscillations is also observed in the large-strength limit of the simple $x \,Z_{N/2}$ perturbation for the VBS-like phase. For the Heisenberg-like phase, the averaging over multiple overlaps between multiple pairs of states may be the reason why similar state-structure effects are not observed, including in Fig. \ref{fig:disordered_sensitivities}. However, some evidence does exist for non-monotonic behavior also in the Heisenberg-like phase for $x \,Z_{N/2-1} Z_{N/2}$ and $x \,Z_{N/2-1}Z_{N/2}Z_{N/2+1}$ perturbations, particularly for perturbation strengths just below values at which the perturbation induces (avoided) crossings involving states of the blind subspace.

\end{document}